\documentclass[aps,twocolumn]{revtex4-1}
\usepackage{graphicx}
\usepackage[justification=justified,width=\linewidth]{caption}
\usepackage{subcaption}
\usepackage{amsmath}
\usepackage{amsfonts}
\usepackage{amsthm}
\usepackage{amssymb}
\usepackage{amsbsy}
\usepackage{wasysym}
\usepackage{bm}
\usepackage{mathrsfs}
\usepackage{color}
\usepackage{times}
\usepackage[resetlabels]{multibib}
\begin{document}
	
	\title{Kinetics of vapor-liquid transition of active matter system under quasi one-dimensional confinement}
	\author {Parameshwaran A and Bhaskar Sen Gupta}
	\email{bhaskar.sengupta@vit.ac.in}
	\affiliation{Department of Physics, School of Advanced Sciences, Vellore Institute of Technology, Vellore, Tamil Nadu - 632014, India}

\begin{abstract}
	We study the kinetics of vapor-liquid phase separation in a quasi one-dimensional confined active matter system using molecular dynamics simulations. Activity is invoked via the Vicsek rule, while passive interaction follows the Lennard-Jones potential. With the system density near the vapor branch, the evolution morphology features disconnected liquid clusters. In the passive limit, coarsening begins with nucleation, followed by an evaporation-condensation growth mechanism, leading to a metastable state without complete phase separation. We aim to understand the impact of Vicsek-like self-propulsion on the structure and growth of these clusters. Our key finding is that Vicsek activity results in a distinct growth mechanism, notably rapid cluster growth and the breakdown of the metastable state through ballistic aggregation. Relevant growth laws are analyzed and explained using appropriate theoretical models.
\end{abstract}

\maketitle

\textit{Introduction.}-- Active matter systems with self-propelled particles often display intriguing clustering when confined, due to the interplay of activity, confinement, and collective behavior \cite{Marchetti,Rao,Ramaswamy,Cates2015}. Real examples span various scales and disciplines. At the microscale, bacterial colonies in microfluidic channels dynamically cluster, responding to environmental cues and geometries \cite{Berg}. At the mesoscale, self-propelled colloids in quasi-2D environments form clusters at liquid-air interfaces, informing materials science \cite{Zottl}. In biophysics, active filaments like microtubules and actin confined in microscale channels organize into dense clusters, influencing cellular mechanics and potential applications in medicine and drug delivery \cite{Mattila}. At the nanoscale, active particles in nanoporous materials exhibit clustering driven by hydrodynamic interactions, enabling control over particle transport with applications in nanofluidics and catalysis, impacting nanotechnology and environmental remediation \cite{Wang}. 

Recently, there has been an upsurge in developing strategies and frameworks to comprehend the active matter systems where self-propelled motions are correlated, leading to the formation of clusters, with a notable emphasis on studying phase transitions and universal behaviors related to scaling properties \cite{Mishra2006,Belmonte,Cates2010,Redner,Wysocki,Mehes,Dey,Mishra2014,Cremer,das2014,trefz2016,das2017pattern1,Chakraborty,Katyal,paul2021,Bera2022,Dittrich}. Experiments have been conducted to investigate the phase behavior and kinetics of phase separation in systems containing active particles \cite{Schwarz,Palacci,Sun,Buttinoni}. Dynamics such as clustering or coarsening involve the evolution of nonequilibrium shapes and the expansion of domains, contrasting with passive systems that rely on factors like symmetry, nature of self propulsion, conservation of order parameters, dimensions of space, and transport mechanisms. Despite similarities in clustering behaviors between active and passive systems, these coarsening characteristics have only been sporadically explored. A compelling but relatively uncharted area is understanding the kinetics of domain growth and structural changes in active matter systems undergoing phase transition, confined within specific geometries. Theoretical challenges to studying such confined systems come from the anisotropic domain evolution, primarily as a limited number of particles can be accommodated along the geometrical constraint. The primary objective of this work is to explore the domain structure formation and growth dynamics in a modeled active matter system under confinement having a certain type of self-propulsion during the vapor-liquid transition.
 
\textit{Model and method.}-- We consider a general model that captures the salient features of an active fluid system embedded in confined geometry. Our system consists of a one component fluid confined within a cylindrical pore. The passive interaction in the system is described by the standard Lennard-Jones (LJ) potential \cite{allen1987,frenkel2002}. The interaction parameter $\epsilon$ and the particle diameter $\sigma$ are chosen as unity. The system described above has a bulk critical temperature $T_c=0.94 \epsilon/k_B$ and critical number density $\rho_c =0.32$ for the vapor-liquid transition in the passive limit \cite{roy2012nucleation}, $k_B$ being the Boltzmann constant. 

While there exist several types of self propulsion models \cite{Redner,Howse,Palacci,tenHagen,Basu,Tailleur}, within the framework aimed at grasping the essence of such clustering phenomena, the Vicsek model \cite{vicsek1995,czirok2000} stands as a cornerstone for the exploration of phase transitions \cite{fisher}. This model emphasizes the tendency of particles to align their movement velocities with the predominant direction observed in their local vicinity. Its applicability extends across a wide range of biological systems that exhibit clustering behaviors. Therefore, we introduce the self-propelled activity through the well known Vicsek-like force as $\Vec{F}^S=F_A \hat{v}_{r_c}$, where $F_A$ is the strength of alignment and $\hat{v}_{r_c}$ being the average velocity direction of the neighbor particles within the interaction range $r_c$. More details on the model and simulation setup are provided in the supplemental material.

The system is studied by molecular dynamics (MD) simulations in the canonical ensemble \cite{allen1987}. To observe the Vicsek activity at an assigned temperature $T$, we use the Langevin thermostat \cite{allen1987,frenkel2002} by solving the following equation,
\begin{equation} \label{eq:1}
\dot{\bm{\Vec{v}}}_i = -\frac{\bm{\Vec{\nabla}} V_i}{m_i}- \zeta\bm{\Vec{v}}_i + \sqrt{\frac{2\zeta K_B T}{m_i}} \, \bm{\Vec{\xi}}_i(t) + \bm{\Vec{F}^S}_i  \bm{;} \hspace{0.2cm}  \Vec{v}_i = \dot{\bm{\Vec{r}}}_i  \hspace{0.2cm}
\end{equation}
Here $m_i$ is the mass of the $i^{th}$ particle, $\Vec{r_i}$ is the position, $\zeta$ is the translational damping coefficient, $V_i$ is the passive interaction potential, and $\Vec{\xi}$ is the  random noise which is delta correlated over space and time as $\langle \Vec{\xi} _{i\mu}(t)\Vec{\xi} _{j\nu}(t') \rangle =\delta_{\mu\nu}\delta_{ij}\delta (t-t')$
where $\mu$ and $\nu$  are the Cartesian components corresponding to the $i^{th}$ and $j^{th}$ particle, $t$ and $t'$ are different times \cite{hansen2008}. The $\Vec{F}^{S}_i$ stands for the Vicsek force.

\begin{figure}[h]
	\centering
	{\includegraphics[width=0.23\textwidth]{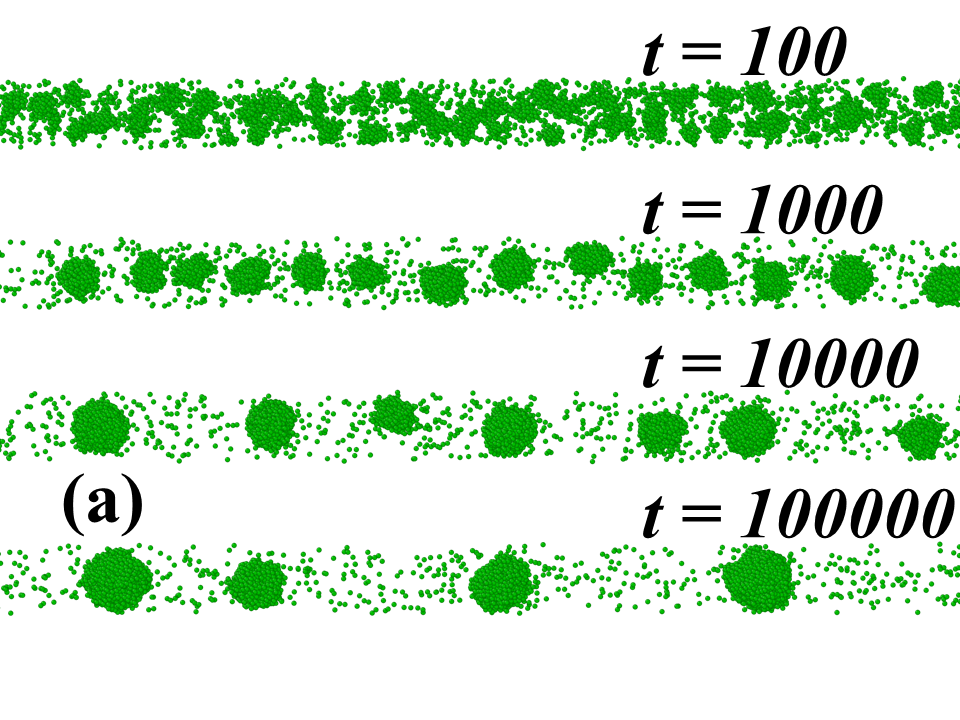}}~~
 \centering
	{\includegraphics[width=0.19\textwidth]{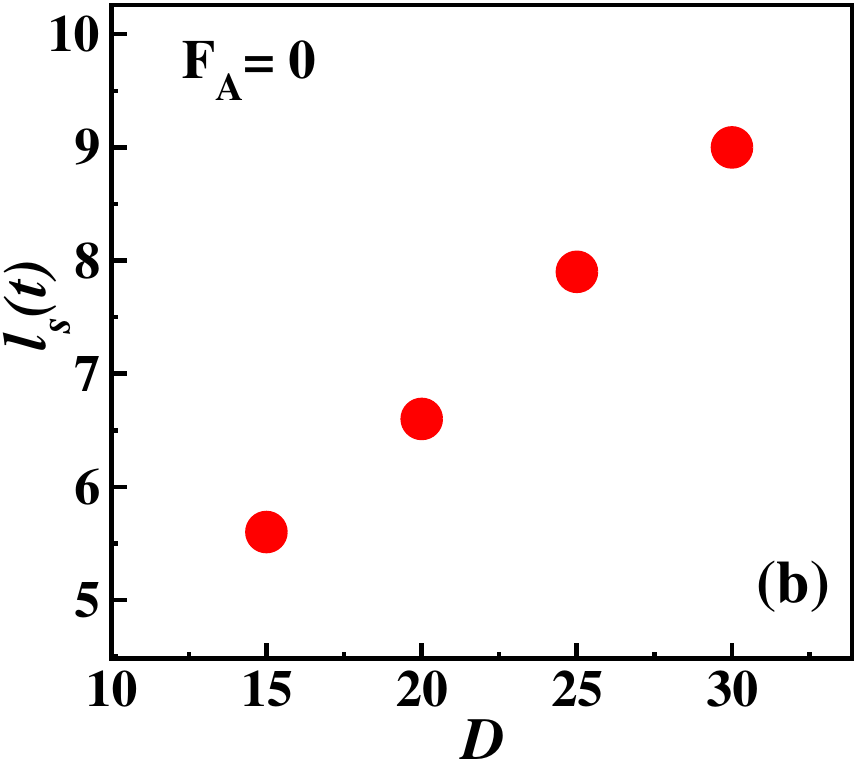}}\\ 
 \centering
	{\includegraphics[width=0.23\textwidth]{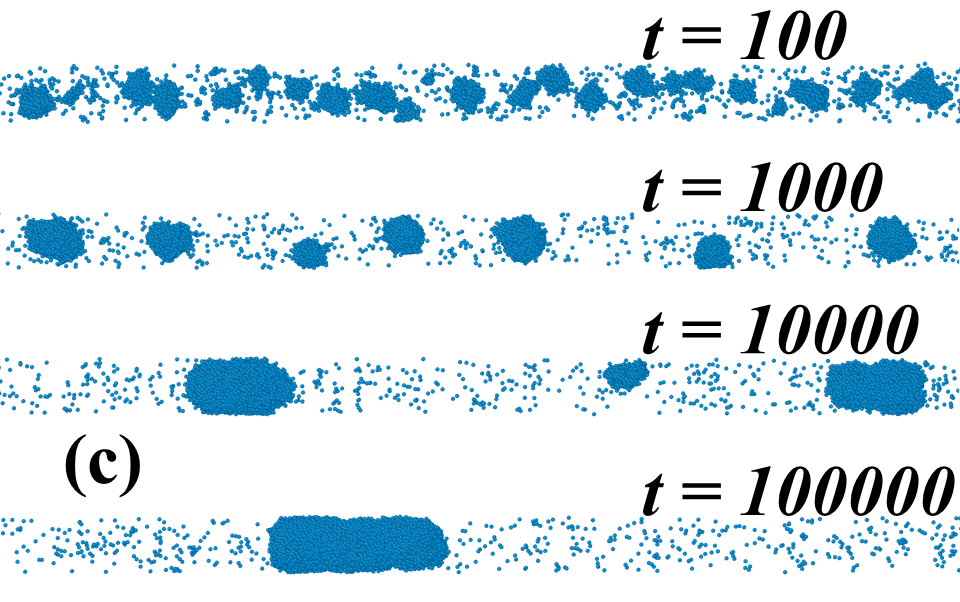}}
	{\includegraphics[width=0.23\textwidth]{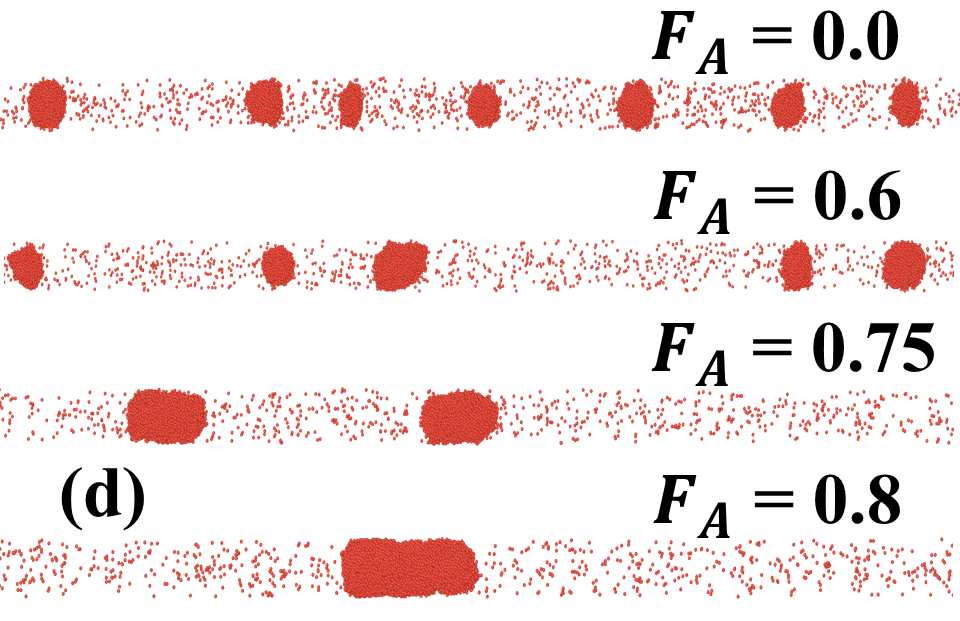}}
	\caption{(a) Evolution snapshots for the passive system. (b) Saturated domain size as a function of the tube diameter. (c) Evolution snapshots for the active system with $F_A=0.8$. (d) The final configurations of our system for different activity values.}
	\label{fig:Of_ac_0}
\end{figure}

The overall density of our system $\rho=0.08$ is chosen close to the vapor branch of the coexistence curve. All the MD runs are carried out using the velocity Verlet algorithm \cite{Verlet} with time step $\Delta t =0.001\tau$, where $\tau = (m\sigma ^2/\epsilon)^{1/2} =1$ is the LJ unit of time. We begin our simulation by equilibrating the system consisting of $N=11400$ classical particles confined in a cylindrical tube at $T=6.0\epsilon/k_B$ using MD simulations to prepare an initial homogeneous configuration in the passive limit ($F_A = 0$). The evolution dynamics of the vapor-liquid phase separation is studied after quenching the system to  $T=0.6\epsilon/k_B$ ($<T_c$), for several values of activity in the range $(0.0$ to $0.8)$. The ensemble average of all the statistical quantities is obtained from 30 independent runs at $T=0.6$ starting from completely different initial configurations.\\
\indent
\textit{Results.}--At the outset, it is worth mentioning that the vapor-liquid phase separation for the passive system in confined geometry has remained unexplored. In Fig.~\ref{fig:Of_ac_0}a the time evolution of the cylindrically confined vapor-liquid system is presented for the passive scenario after the quench at $T=0.6\epsilon/k_B$. Due to the overall density close to the vapor branch of the coexistence curve, phase separation initiates via nucleation, followed by the growth of spherical liquid droplets, similar to the bulk system \cite{roy2012nucleation,bindermidya2017droplet}. Here, once the droplets form, they effectively remain static under the influence of the Langevin thermostat and grow in size through the evaporation-condensation mechanism, as proposed by Lifshitz and Slyozov (LS) \cite{lifshitz1961}. In this process, particles detach from smaller droplets and undergo diffusive deposition on to larger droplets. The droplets maintain a spherical shape to optimize the interfacial free energy \cite{bray2002}. Ultimately, periodic spherical domains form along the axis of the cylinder, prompting the system to stabilize into a metastable state. Therefore, unlike the bulk system, a complete macroscopic phase separation is never achieved. The reason is attributed to the lack of interaction between neighboring domain interfaces when they are separated beyond a certain distance. This scenario is found to be robust, and domain size is insensitive to the length of the cylinder but varies linearly with the tube diameter, as demonstrated in Fig.~\ref{fig:Of_ac_0}b 

However, the introduction of Vicsek activity adds a layer of complexity to the growth phenomenon. When the activity is turned on, the coarsening process accelerates significantly with the strength ($F_A$) of self propulsion. At the later stage, periodic spherical liquid droplets coalesce to form plug-shaped domains that subsequently merge and continue to grow. This dynamic process ultimately disrupts the metastable state, paving the way for complete phase separation. This is demonstrated in Fig.~\ref{fig:Of_ac_0}c by displaying the time evolution of the vapor-liquid phase separation for a relatively high activity case of $F_A = 0.8$ where the difference in structural change of the active system with respect to the passive limit is best realized within our simulation time frame. In Fig.~\ref{fig:Of_ac_0}d we show the final configurations of our system corresponding to different activity values for the maximum MD time in our simulation. The accelerated domain growth via the coalescence of droplets with increasing activity is evident from these results. Further discussion on this is provided later.
\begin{figure}[h]
	\centering
	{\includegraphics[width=0.22\textwidth]{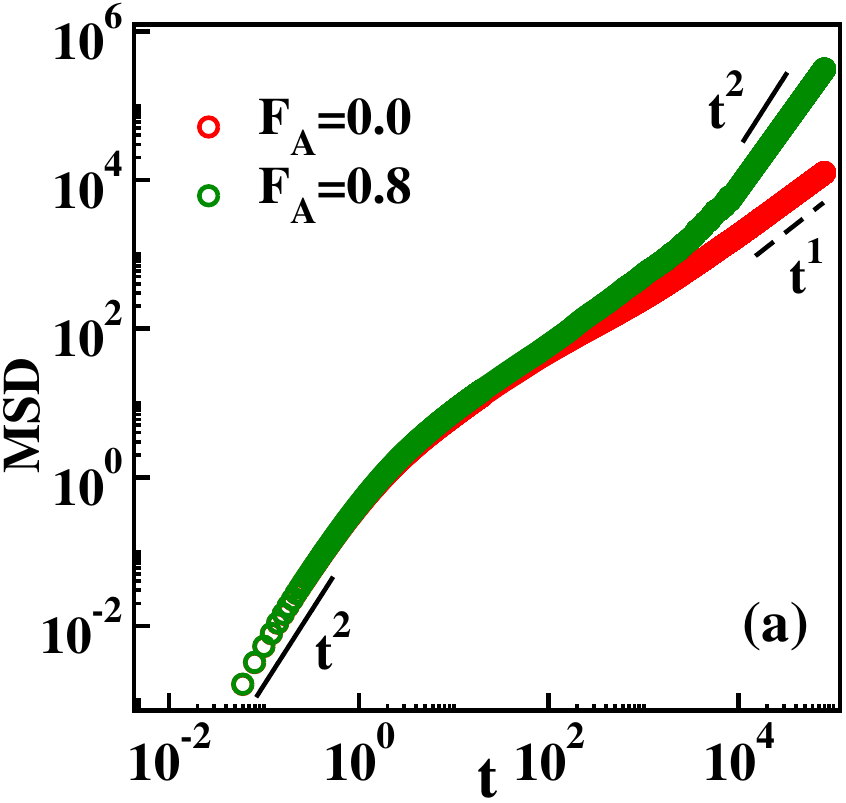}}
	{\includegraphics[width=0.24\textwidth]{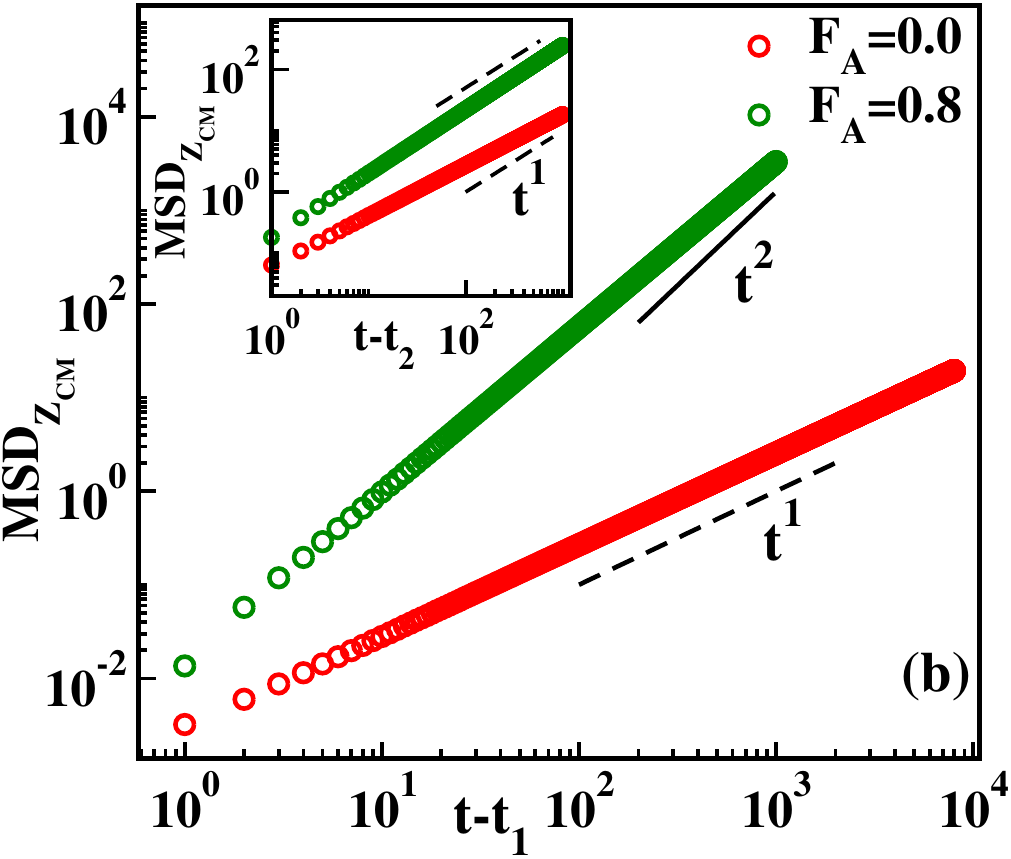}}
	\captionsetup{justification=raggedright,singlelinecheck=false}
	\caption{ (a) The MSD of a single particle corresponding to activity $F_A=0$ and $F_A=0.8$. (b) Later time MSD of the CM of liquid clusters for activity $F_A=0$ and $F_A=0.8$ is shown in the main figure. Inset shows the same at the initial time. Here $t_1=10000$ and $t_2=1000$ The doted lines indicate diffusive regime ($t^1$) and solid lines indicate ballistic regime ($t^2$).   }
	\label{fig:O_Msd}
\end{figure}

To understand the coarsening mechanism, we compute the mean-squared displacement (MSD) for a single particle and the center of mass (CM) of the clusters. The expression for the later is given by $\mathrm{MSD}_{cm}= \langle ( \Vec{R_{cm}}(t)-\Vec{R_{cm}}(t_0) )^2  \rangle$. Here $\Vec{R_{cm}}(t)$ is the CM of a cluster at time $t$ given by \cite{hansen2008},
\begin{equation} \label{eq:2}
	\Vec{R_{cm}} = \frac{1}{M_c} \sum_{i=0}^{N_c} m_i \Vec{r_i}
\end{equation} 
$M_c$ and $N_c$ are the mass and number of particles of the cluster. $t_0$ represents the initial time of measurement. A comparative picture for the active and passive cases is presented in Fig.~\ref{fig:O_Msd}a for the single particle MSD. The early time ballistic behavior crosses over to the standard diffusive regime for both cases. At the intermediate time, the active system starts to deviate towards a super-diffusive regime. Finally, a robust ballistic behavior is observed due to the Vicsek activity. Fig.~\ref{fig:O_Msd}b demonstrates the average MSD for the clusters. From the results, it is evident that the value of MSD in the passive case is two orders of magnitude lower than in the active system, and the clusters for the earlier system effectively remain stationary. Contrary to this, in the presence of activity, the clusters move in a diffusive manner. A transition from diffusive to ballistic motion is observed at the later stage. Since coherency is a fundamental aspect of the Vicsek model, the directed motion of particles facilitates this outcome. This allows the clusters to move and merge along the cylindrical axis via the ballistic cluster coalescence mechanism.

To quantify the domain growth dynamics, we resort to the so-called two-point equal time correlation function $C(z,t)$ \cite{bray2002,daniya} along the length of the cylindrical geometry given as,
\begin{equation} \label{eq:3}
	C(z,t)= \langle \psi(0,t)\psi(z,t) \rangle  -\langle \psi(0,t)\rangle \langle \psi(z,t)\rangle
\end{equation}
Here, the angular brackets denote statistical averaging over different initial configurations, and $\psi$ represents the order parameter. To compute $\psi(z,t)$, we partition the entire cylinder into cubic boxes of size $(2\sigma)^3$. Subsequently, the order parameter associated with each box is assigned the value $\psi(z,t) = +1$ if the local number density exceeds $\rho_c$ and $\psi(z,t) = -1$ otherwise \cite{daniyaG}.
\begin{figure}[h]
	\centering
	{\includegraphics[width=0.23\textwidth]{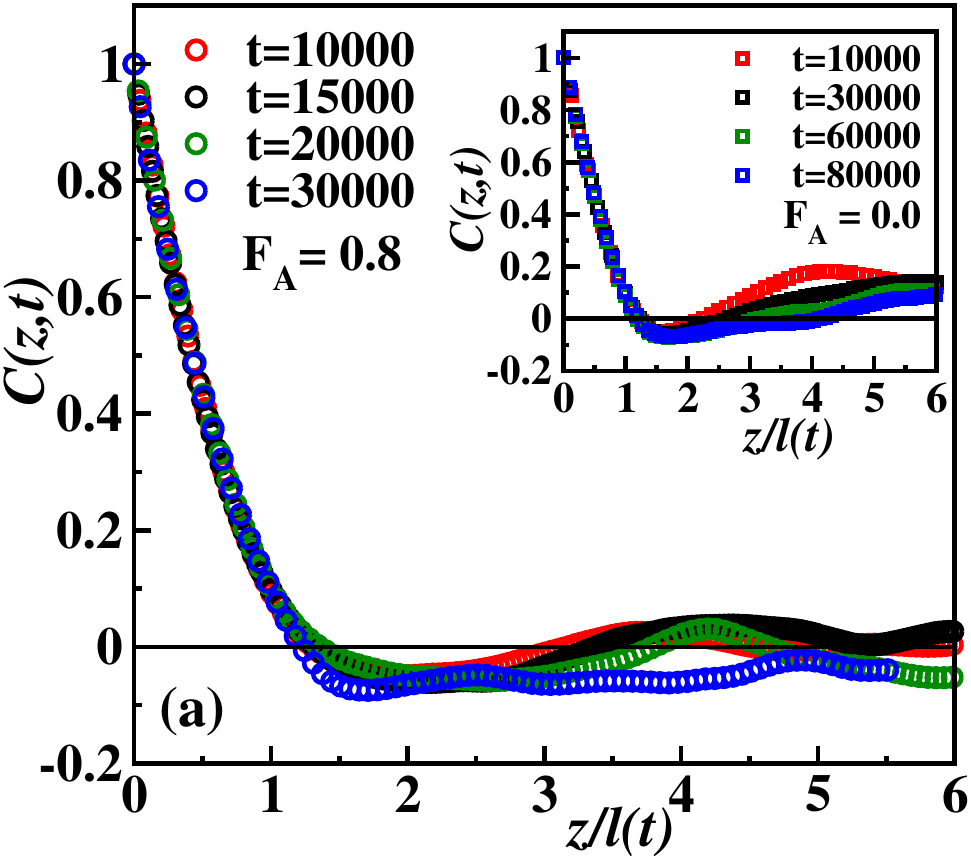}}
	{\includegraphics[width=0.235\textwidth]{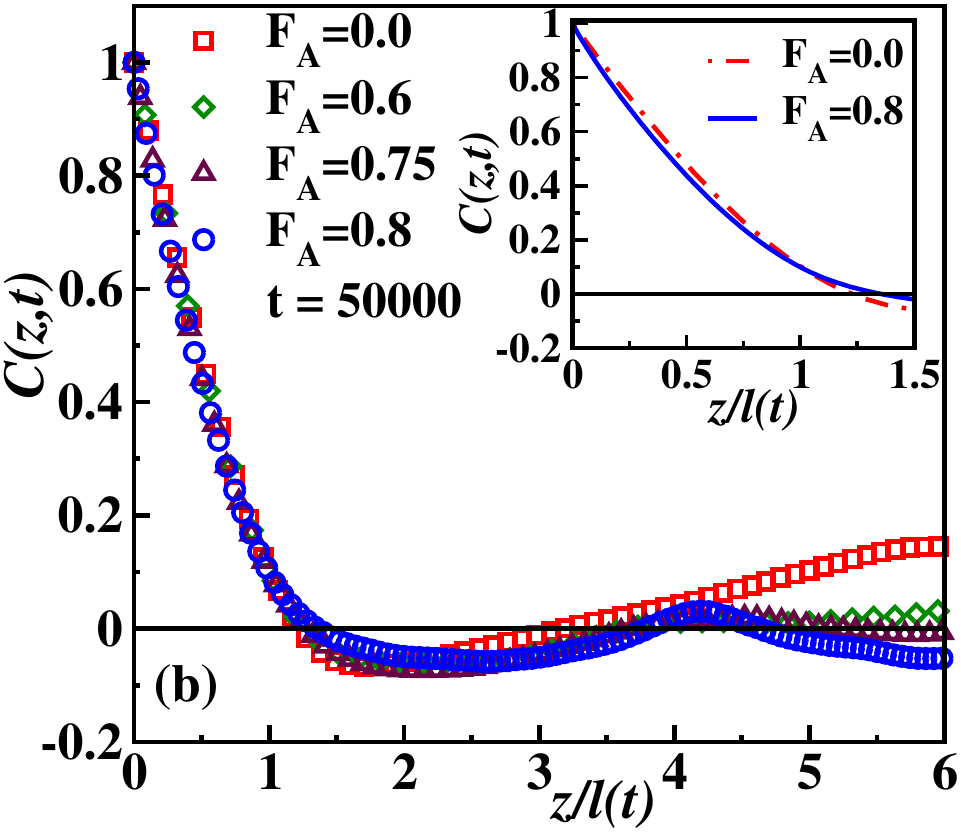}}
	\captionsetup{justification=raggedright,singlelinecheck=false}
	\caption{(a) Scaling plots of the correlation function $C(z,t)$ for the activity (main) and passive system (inset) for different time values. In (b) we show the same scaling plots for different activity levels at a fixed time (main), and the horizontal axis limited to $z/\ell =1.5$ for enhanced clarity (inset).}
	\label{fig:O_cor1}
\end{figure}
For the patterns that are statistically self-similar, $C(z,t)$ exhibits the scaling property as $C(z,t)\equiv \tilde{C}(z/\ell(t))$, where $\tilde{C}$ is a master scaling function independent of time \cite{bray2002,ahmed}. This scaling law facilitates the definition of the relevant time-dependent characteristic length scale $\ell(t)$, representing the average domain size. In this paper, we use the first 0.1 crossing of $C(z,t)$ as a measure of $\ell(t)$.  

In Fig.~\ref{fig:O_cor1}(a)  we present the scaling plot of $C(z,t)$ with $z/\ell$, for both the active and passive case, utilizing various time data. A notable data collapse suggests that the evolving structures maintain self-similarity throughout the observed time span. Likewise, in Fig.~\ref{fig:O_cor1}(b) a scaling plot of $C(z,t)$ with $z/\ell$ is presented, showcasing variations with different activity values. For $z/\ell<1$, the data collapse is quite satisfactory with minor discrepancies, indicating the differences in domain morphology. This results from the influence of Vicsek activity in the structure formation, which tries to keep the particles within the cluster, especially those in the peripheral regions, moving in relation to the center of mass and making domains more compact and less fractal than passive cases. This argument is substantiated by computing the density profile of the liquid phase in the active and passive cases and is provided in the supplementary material.  
\begin{figure}[h]
	\centering
	{\includegraphics[width=0.225\textwidth]{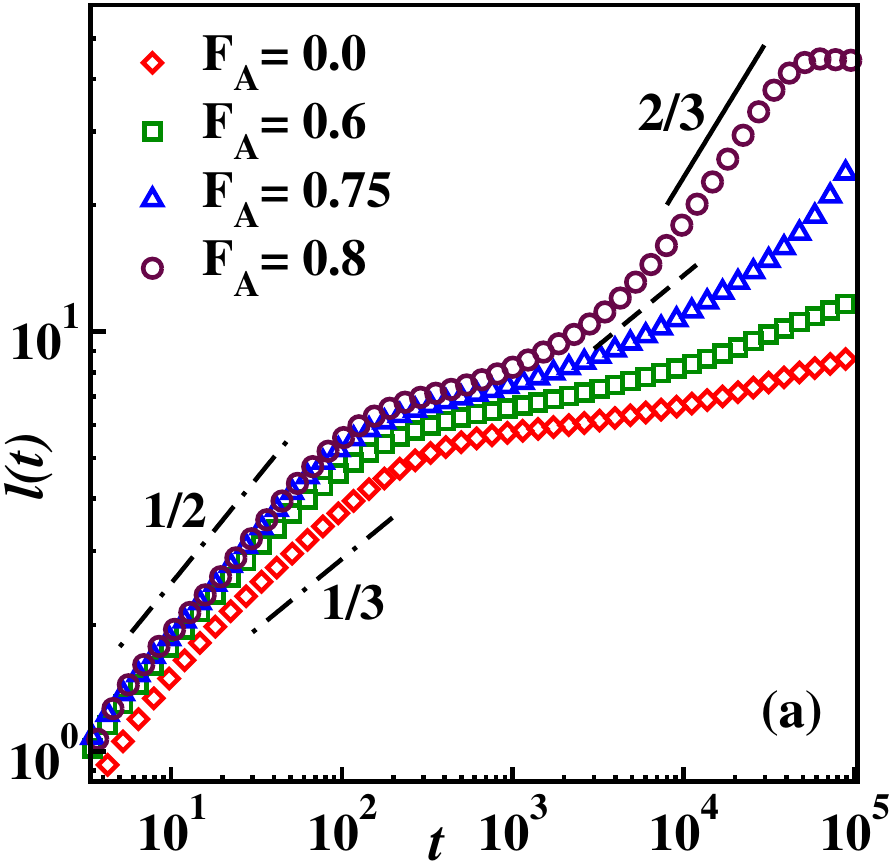}}
	{\includegraphics[width=0.24\textwidth]{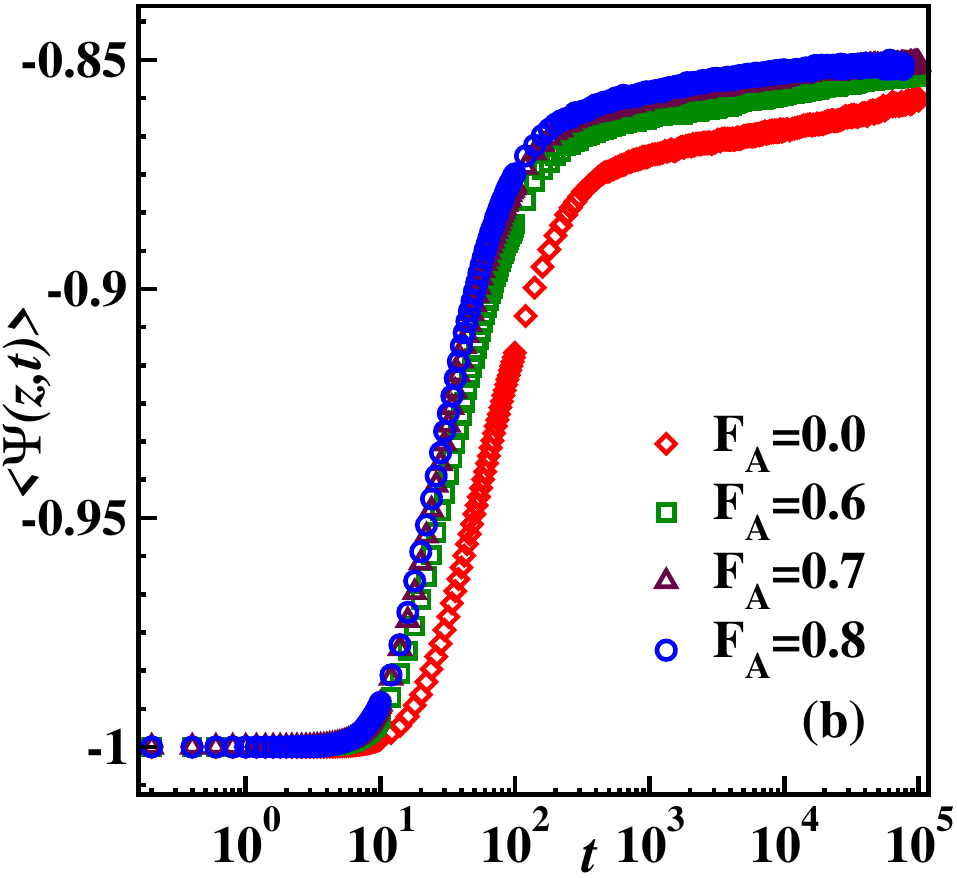}}
	\caption{(a) Time dependence of the characteristic length-scale $\ell(t)$ for different values of activity. (b) Plot of the average order parameter for systems with different activity as a function time.}
	\label{fig:O_domain}
\end{figure}

The evolution morphology is best realized in terms of the domain size $\ell(t)$. In Fig.~\ref{fig:O_domain}a we show the time-dependence of the $\ell(t)$ against $t$ for the liquid domain on a log-log scale for various activity levels computed from the $C(z,t)$ following the method mentioned above. It is conspicuous that for both passive and active systems, domain growth follows the power-law $\ell(t) \sim t^{\alpha}$. In the passive case, initially the data aligns with $\alpha=1/2$. This is in accordance with the non conserved order parameter dynamics \cite{bray2002} in the initial stage and is illustrated in Fig.~\ref{fig:O_domain}b. Subsequently, the data aligns consistently with $\alpha=1/3$, reflecting the Lifshitz and Slyozov (LS) evaporation-condensation growth mechanism \cite{lifshitz1961}. Eventually, the phase separation comes to a halt when the domain size becomes comparable to the diameter of the cylinder. The reason is attributed to the lack of interaction between the droplets, and therefore, the droplet diffusion is negligibly small under the influence of stochastic Langevin thermostat as demonstrated via their MSD. 

Contrary to this, we observe a noteworthy increase in the growth exponent $\alpha$ at late time when the activity is raised from $0$ to $0.8$. For the active case, there is a brief initial period of non-conserved dynamics where $\alpha$ is close to $1/2$. This is followed by the Binder and Stauffer (BS) Brownian coalescence mechanisms \cite{binder1974,binder1977}. In this regime, self-propulsion begins to take effect, and unlike the passive case the droplets exhibit diffusive motion, as demonstrated in Fig.~\ref{fig:O_Msd} via MSD. 
\begin{figure}[h]
	\centering
	{\includegraphics[width=0.23\textwidth]{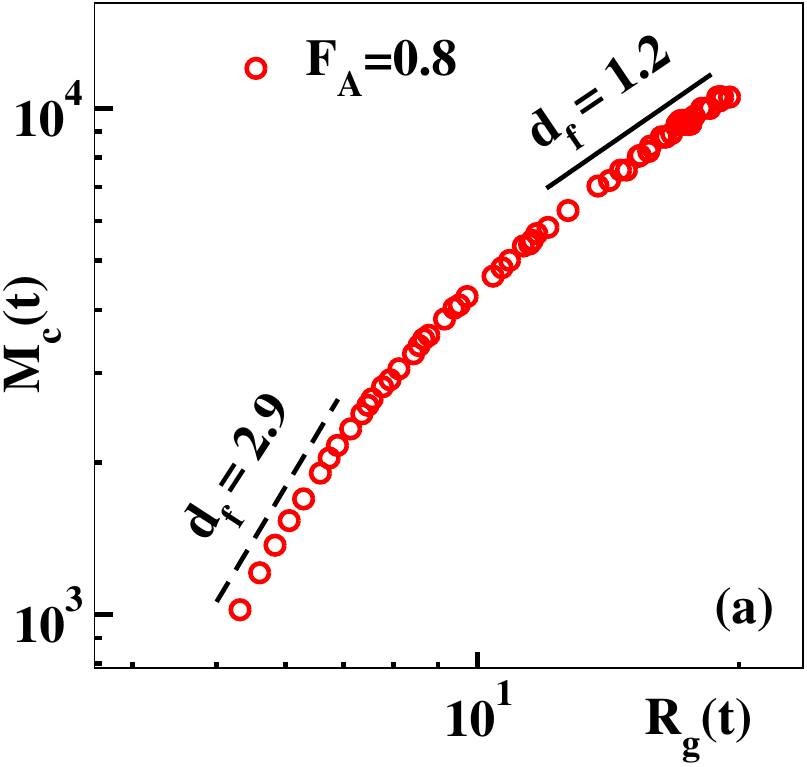}}
	{\includegraphics[width=0.23\textwidth]{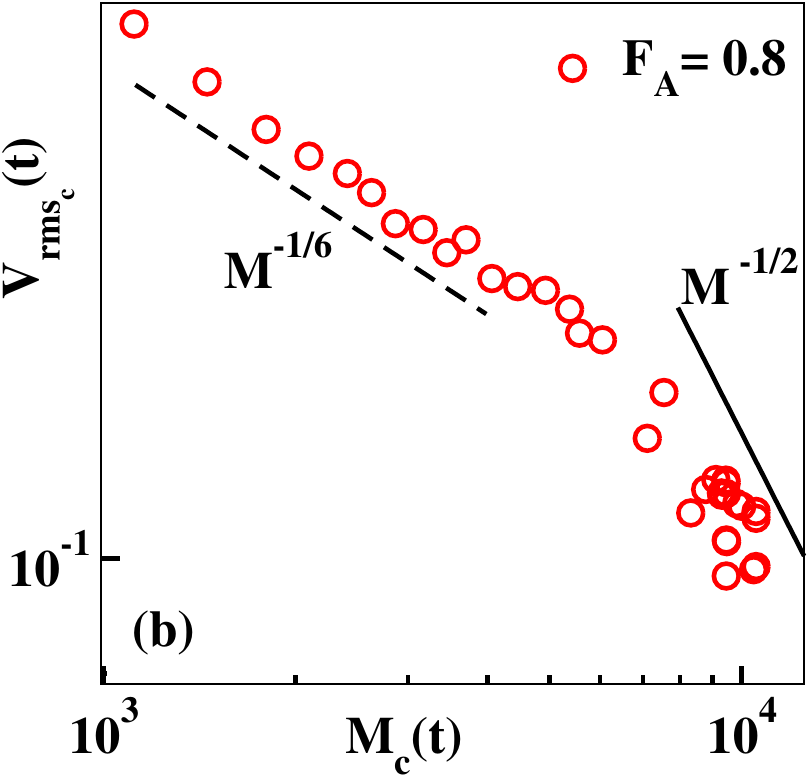}}
	\caption{ (a) The average mass $M_c$ vs average radius of gyration $R_g$ of the liquid clusters plotted on a log-log scale. (b) Log-log plot of the RMS velocity $v_{rms}$ vs the average mass $M_c$ of the clusters. The solid lines represent power-law behavior.}
	\label{fig:O_m_vrms}
\end{figure}
Further discussion on this is added in the supplemental material. For the BS mechanism, the change in droplet density \( \rho_n \)  with time \( t \) is governed by the differential equation \cite{binder1974,binder1977}
\begin{equation}\label{eq:4}
\frac{d\rho_n}{dt} = -C\rho_n^2
\end{equation}
 where \( C \) is a constant related to the Stokes-Einstein-Sutherland relation \cite{hansen2008,onuki2002}. The solution of this equation yields $\rho_n(t) \propto 1/t$. Introducing the relationship \( \rho_n \propto 1/\ell^{d} \), where \( \ell \) represents the average cluster size at time \( t \) in \( d \)-dimension, we get
\begin{equation}\label{eq:5}
\ell(t)\propto t^{1/d}\propto t^\alpha
\end{equation}

To estimate the $d$ in confined geometry, we compute the fractal dimension $d_f$ \cite{vicsek1992,matsushita1994} of the evolving clusters of the liquid using the relation $M_c \propto R_g^{d_f}$. Here $M_c$ is the average mass of the clusters and  $R_g$ is the average radius of gyration given by,
\begin{equation} \label{eq:6}
	R_g = \left< \left(\frac{1}{N_c}\sum_{i=0}^{N_c}(\Vec{r_i}-\Vec{R_{cm}} )^2\right)^{1/2} \right> 
\end{equation}
The result is displayed in Fig.~\ref{fig:O_m_vrms}a. In the early stage, the impact of confinement on the growing droplets is minimal, leading to $d_f \approx 3$ as illustrated in Fig.~\ref{fig:O_m_vrms}a, resulting in $\alpha \approx 1/3$. 

In the later stage, the growth exponent deviates from that of the passive case. For the intermediate activity strength, the growth rate enhances gradually, depending on the competition between active and passive interactions. For the choice of higher activity of $F_A=0.8$, self-propulsion dominates over the passive interaction, and the saturation of length scale is achieved within our maximum simulation time. In this case, the growth aligns with $\alpha \approx 2/3$. This exponent can be elucidated by considering the theory of ballistic coalescence mechanisms \cite{carnevale1990,trizac2003,paul2017ballistic}. Here the change of droplet density $\rho_n$ over time $t$ is governed by the differential equation 
\begin{equation} \label{eq:7}
	\frac{d\rho_n}{dt} = -\rho_n\nu_c
\end{equation} 
where $\nu_c$ is the average collision time. For the given dimension $d$ 
\begin{align}\label{eq:8}
\nu_c &= v_{rms}/\lambda \quad\bm{;} \quad \lambda \propto 1/\{\rho_n \ell^{(d-1)}\}
\end{align} 
Here $v_{rms}$ and $\lambda$ are the root-mean-squared velocity and mean free path of the clusters respectively. From (\ref{eq:7}) and (\ref{eq:8}) we get, 
\begin{equation} \label{eq:9}
	\frac{d\rho_n}{dt} \propto -\rho_n^{(d+1)/{d}} v_{rms}
\end{equation}
Under the influence of Langevin thermostat the clusters can be assumed to be uncorrelated \cite{carnevale1990}, and the kinetic energy follows the Boltzmann distribution, whereby \( v_{rms} \propto M_c^{-1/2} \propto \rho_n^{1/2} \). This relationship is illustrated in Fig.~\ref{fig:O_m_vrms}b. The low droplet density at later stages results in noisy data. A detailed explanation is provided in the supplemental material. Subsequently, we get 
\begin{equation} \label{eq:10}
	\frac{d\rho_n}{dt} \propto -\rho_n^{(3d+2)/{2d}}
\end{equation}
The solution of this equation gives, 
\begin{equation} \label{eq:11}
	\rho_n(t) \propto t^{-{2d}/(d+2) } \quad \text{and} \quad  \ell(t) \propto t^{{2}/(d+2)}  \propto t^\alpha 
\end{equation}
Therefore, we get \(\alpha= {{2}/(d_f+2)}\). For enhanced precision of dimensionality, we compute the fractal dimension ($d_f$). In Fig.~\ref{fig:O_m_vrms}a, it is evident that the late time growth structure follows $d_f \approx 1$, which results in $\alpha \approx 2/3$. The growth plateau observed at later times for $F_A=0.8$ is attributed to the finite-size effect. 

\textit{Conclusions.}-- This study presents a quantitative picture of the influence of Vicsek like self-propulsion on various aspects of kinetics of vapor liquid phase separation of a modeled active matter system exhibiting discontinuous morphology under confined geometry. The basic characteristics of phase ordering dynamics apply to both passive and active systems. Self-similar domain growth is observed through scaling in the two-point equal-time functions. This is intriguing as an active matter system advances towards a steady state, as opposed to a passive system that moves towards equilibrium. Nevertheless, there are striking differences in the coarsening dynamics between the two systems. In the passive case, we have identified two power law regimes, the early stage nucleation followed by Lifshitz-Slyozov diffusive mechanism with exponent $\alpha =1/3$. At the late time, the clusters appeared to be frozen, and the system attains a metastable state.  For the active system, the domains grow via the Binder and Stauffer Brownian coalescence mechanism after the nucleation. Finally, at late time, a hydrodynamic like behavior in the active matter domain growth is observed due to the fast coherent ballistic motion of the particles, whereby a complete phase separation is achieved. We have quantified the growth law as $\alpha=2/3$ and explained it by obtaining accurate knowledge on the mass dependence of the velocities of the clusters. It will be interesting to extend our work in other active matter models to investigate the existence of universal behavior in the phase separation dynamics in confined geometry.

\textit{Acknowledgements.}--B. Sen Gupta acknowledges Science and Engineering Research Board (SERB), Department of Science and Technology (DST), Government of India (no. CRG/2022/009343) for financial support. Parameshwaran A. acknowledges DST-SERB, India for doctoral fellowship.

\clearpage
\onecolumngrid
\section*{Supplementary Material}

\section{Simulation Setup}
We consider a single-component classical $N$ particle system that undergoes vapor-liquid phase separation confined inside cylindrical pore. The passive interaction between a pair of particles $i$ and $j$, separated by a distance $r_{ij}=|r_i - r_j|$, is described by the standard Lennard-Jones (LJ) potential given by,
\begin{equation}
	U(r_{ij})=4\epsilon\left[\left(\frac{\sigma}{ r_{ij}}\right)^{12}- \left(\frac{\sigma}{r_{ij}}\right)^6\right] \label{eq1} .
\end{equation}
Here, $\sigma$ and $\epsilon$ refer to the diameter of the particle and the strength of interaction, respectively. For the sake of convenience, we set $m$ and $k_B$ to unity. To improve computation speed, the potential is truncated at $r_c=2.5\sigma$. However, the insertion of a cut-off radius can cause discontinuities in the force and potential. To address this issue, the potential is modified as \cite{Bhattacharyya1}
\begin{equation}
	V(r_{ij}) = \begin{cases}
		\text{if } (r_{ij} < r_c); \vspace{0.05cm}\\
		U(r_{ij}) - U(r_c) - (r_{ij} - r_c) \left(\frac{dU}{dr_{ij}}\right)_{r_{ij}=r_c}, \vspace{0.2cm}\\
		
		\text{if } (r_{ij} \geq r_c) ; \hspace{1cm}0 .
	\end{cases}
\end{equation}

\begin{figure}[h]
	\centering
	{\includegraphics[width=0.48\textwidth]{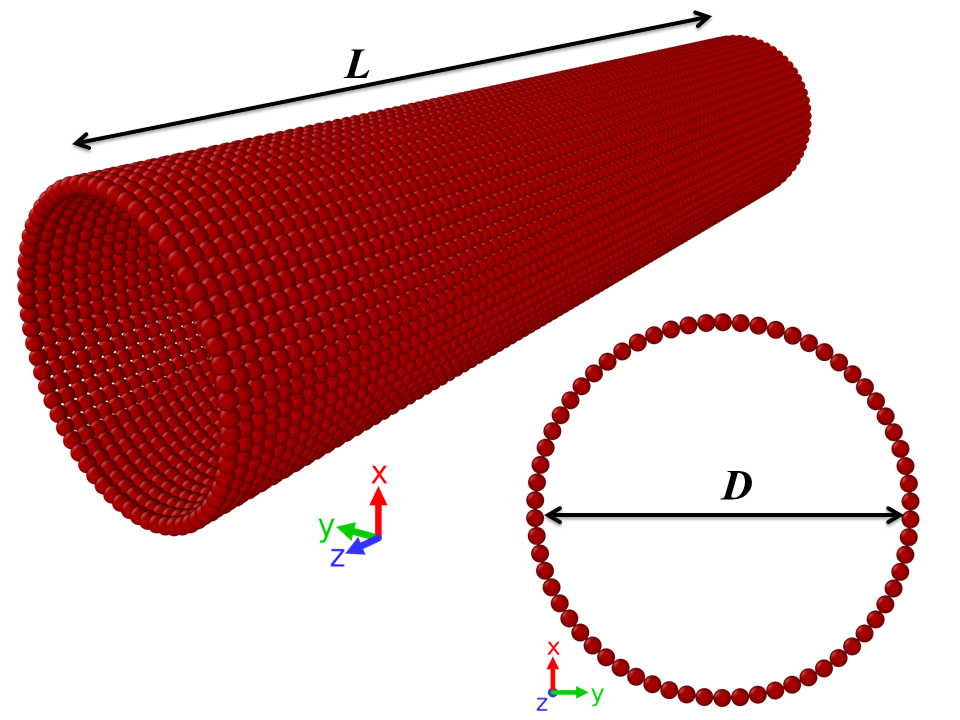}}
	\caption{Schematic diagram of the cylindrical host structure with the axis oriented along z-direction.}
	\label{fig:cylinder}
\end{figure}

The structure of the cylindrical pore is shown in Fig.~\ref{fig:cylinder}. It is constructed using particles with a number density of $\rho_w = 1$, and  have the same diameter $\sigma$ as the fluid particles. The wall particles are kept frozen throughout the simulation. The  length of the cylinder $L=500$  is considered to be much larger than its diameter $D=20$. Here the length is expressed in units of $\sigma$.
The wall particles interact with the fluid particles via LJ potential, the interaction range being truncated at $2^{\frac{1}{6}}\sigma$. Periodic boundary condition is applied along the length of the cylinder. This allows us to create the illusion of an infinitely long system.

The self-propelled activity is introduced through the Vicsek-like rule as \cite{vicsek19951,czirok20001},
\begin{equation}
	\Vec{F_i}^S=F_A \hat{v}_{r_c},
\end{equation}
where $F_A$ is the strength of alignment interaction and $ \hat{v}_{r_c}$ is the average velocity direction of the neighbor particles of $i$, contained within the sphere of radius $r_c = 2.5$, given by
\begin{equation}
	\bm{\hat{v}_{r_c}} = \frac{\sum_j \Vec{v_j} }{|\sum_j \Vec{v_j} |} ,\
\end{equation}
Therefore, the final velocity of each particle can be written as a superposition of the passive as well as the active part,
\begin{equation} \label{eq2}
	\Vec{v_i}(t+\Delta t)= \Vec{v_i}^{pas}(t+\Delta t) +\frac{\Vec{F_i}^S}{m_i}\Delta t 
\end{equation}
However, introducing self propelling force in the above form not only alters the direction of velocity but also the magnitude. Consequently, the system temperature rises and stabilizes at much higher levels than anticipated. This significantly impacts the phase behavior of the system, as temperature plays a critical role. To address this issue, we modify equation (\ref{eq2}) as \cite{das2017pattern1}
\begin{equation}
	\Vec{v_i}(t+\Delta t)=|\Vec{v_i}^{pas}(t+\Delta t)| \hat{n} ,
\end{equation}
where $\hat{n}$ is given by
\begin{equation}
	\hat{n}=\Bigl(\Vec{v_i}^{pas}(t+\Delta t) +\frac{\Vec{F_i}^S}{m_i}\Delta t\Bigl) \Bigl/ \Bigl( |\Vec{v_i}^{pas}(t+\Delta t) +\frac{\Vec{F_i}^S}{m_i}\Delta t|\Bigl)
\end{equation}
Thus, by following this method, the self-propelling force applied to each particle alters only the direction of motion, similar to the original Vicsek model. This approach enables us to more accurately control the system temperature and study the phase behavior.

\section{Domain Structure Analysis}
\subsection{Static Structure Factor}
To gain insight into the morphology of domain boundaries, we compute the structure factor \cite{Bhattacharyya2,Bhattacharyya3}, the Fourier transform of the correlation function $C(z,t)$, given as
\begin{equation}
	S(k_z,t)=\int e^{ikz}C(z,t) dz
\end{equation}
where $k_z$ is the wave vector. For self similar patterns, $S(k_z,t)$ has the scaling form
\begin{equation}
	S(k_z,t)\equiv \ell^d \tilde{S}(k_z\ell(t)),
\end{equation}
Here $\tilde{S}$ is a time-independent scaling function. In $d$-dimension the decaying part of the tail of $S(k_z,t)$ shows the Porod law behavior as $S(k_z,t) \sim k^{-(d+1)}$ \cite{bray120021}. 
\begin{figure}[h]
	\centering
	{\includegraphics[width=0.4\textwidth]{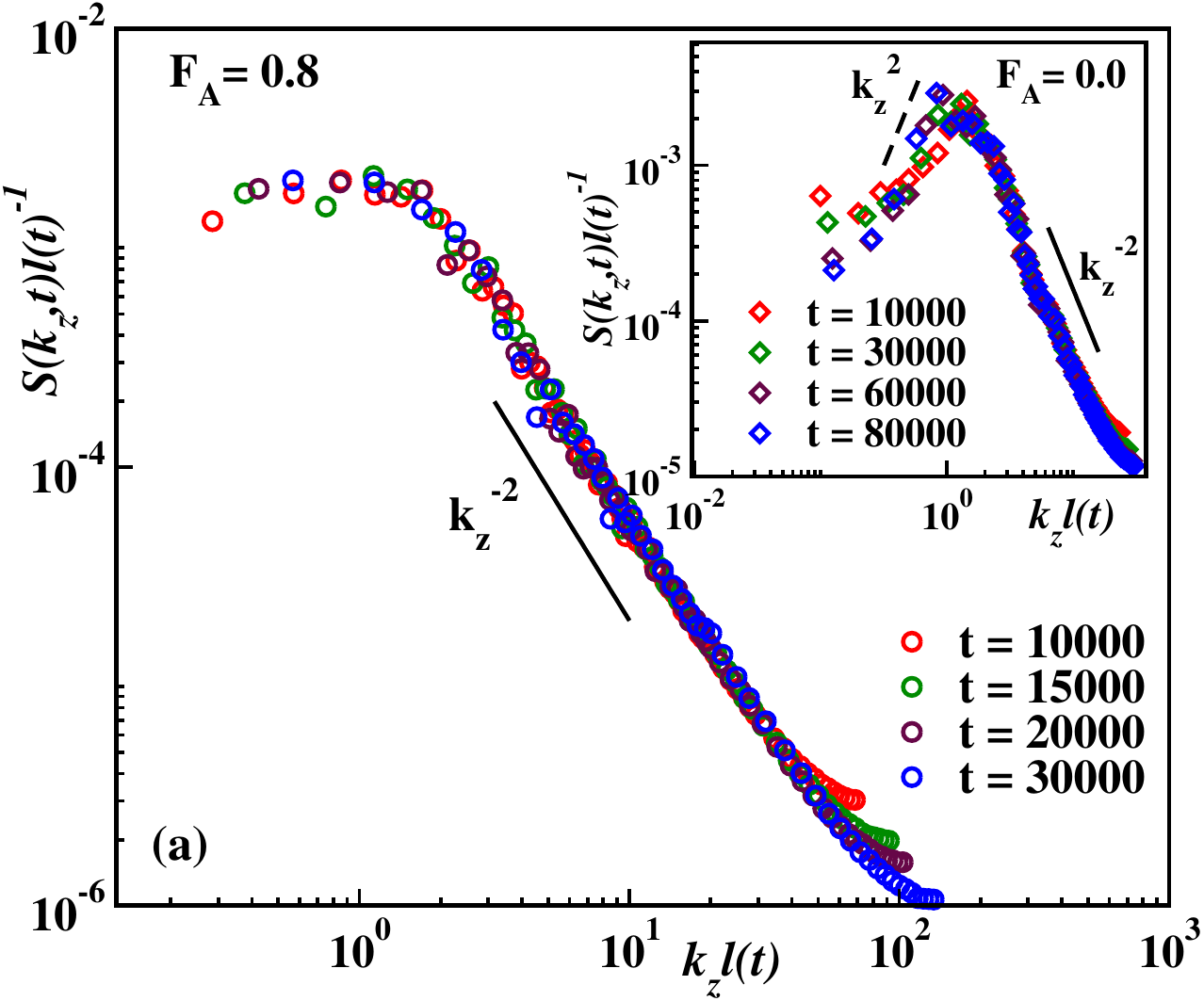}}
	{\includegraphics[width=0.4\textwidth]{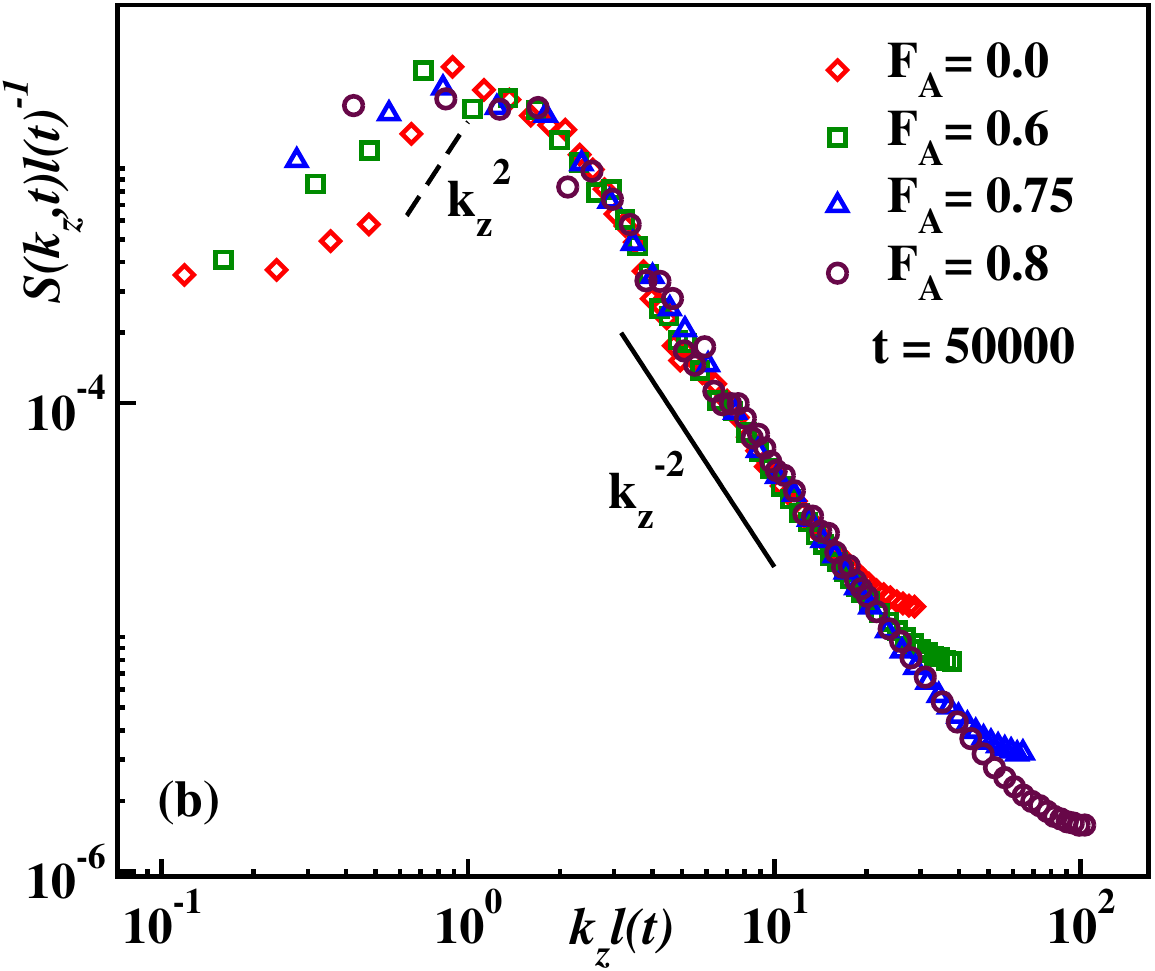}}
	\captionsetup{justification=raggedright,singlelinecheck=false}
	\caption{(a) Scaling plots of the structure factor for the active system. The inset represents the same in the passive limit (b) Scaling plots of the structure factor for different activity levels at a fixed time.}
	\label{fig:O_struct1}
\end{figure}
In  Fig.~\ref{fig:O_struct1}(a)  we show the log-log scaling plot of $S(k_z,t)\ell^{-1}$ vs $k_z\ell(t)$ for the passive and active case ($F_A=0.8$), using different time data. Likewise, in Fig.~\ref{fig:O_struct1}(b) we present scaling plot for different activity values. In all cases, a satisfactory data collapse is evident. At large $k_z$ limit, the decaying part of the tail exhibits a power law $S(k_z,t) \sim k_z^{-2}$, showing the behavior of the Porod law and supporting the concept of a sharp interface \cite{Shrivastav}.
\subsection{Density profile of domains}
\begin{figure}[h]
	\centering
	{\includegraphics[width=0.4\textwidth]{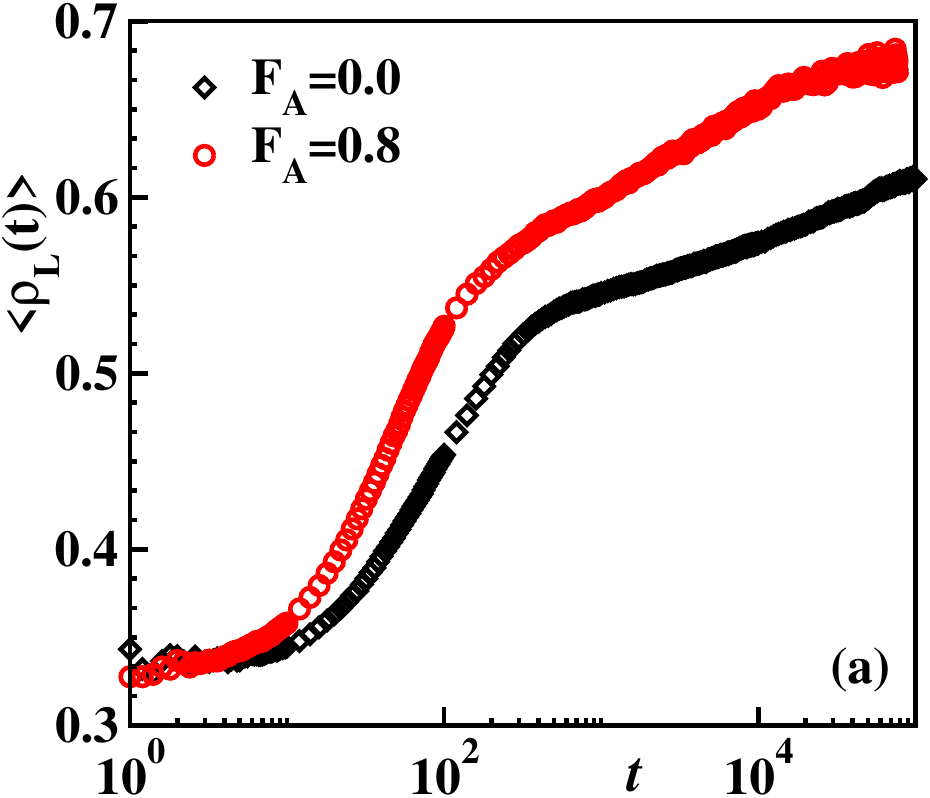}}
	{\includegraphics[width=0.4\textwidth]{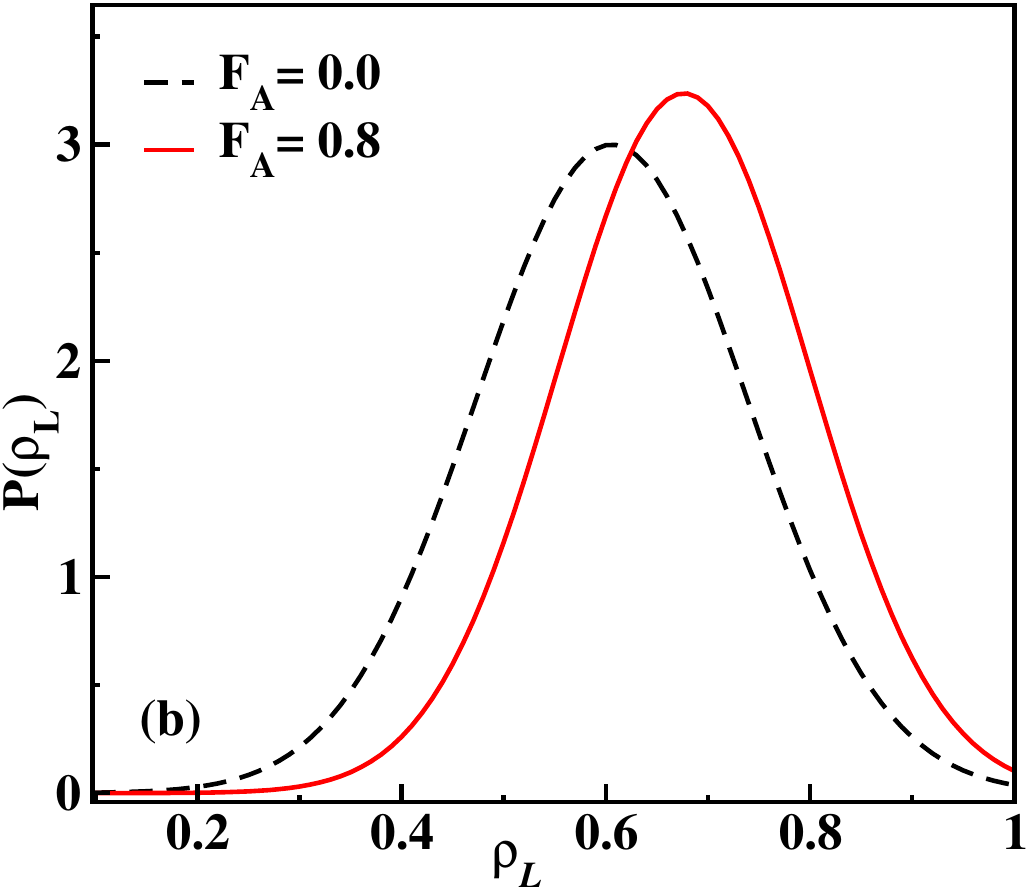}}
	\captionsetup{justification=raggedright,singlelinecheck=false}
	\caption{(a) The variation of the average density of the liquid domains with time. (b) The density distribution of the fully separated liquid domain. Results for both the active and passive systems are presented.}
	\label{fig:density}
\end{figure}

Since coherency is an intrinsic aspect of the Vicsek model, it is expected that the separated phases are affected by the activity. For better understanding, we compute the density of the liquid domains for the active and passive system. In Fig.~\ref{fig:density}a we show the variation of the average density of the liquid domain with time for both systems. 
The results clearly demonstrate that the liquid domains compactify with time under the influence of activity. In Fig.~\ref{fig:density}b we present the distribution of average density  $P(\rho_l)$ of the completely separated liquid phase for the active and passive system at long time. The peak position of the distribution represents the average density. The results substantiate the fact that the activity results in tightly packed liquid domain.

\section{Diffusive motion of clusters}
If the liquid clusters exhibit random motion within the system, they can be modeled as undergoing Brownian motion, making the mean squared displacement (MSD) a suitable measure for quantifying their dynamics. The MSD for a cluster can be expressed as:
\begin{equation}
	\langle r^2  \rangle = 2d D_t t
\end{equation}
where \( \langle r^2 \rangle \) represents the average of squared displacement over time, \( d \) is the dimension of the system, \( D_t \) is the translational diffusion coefficient, and \( t \) is time. The root mean square velocity (\( v_{rms} \)) of the clusters can be derived from the MSD as follows
\begin{equation} \label{eq:vrms}
	v_{rms} =  \frac{\langle r^2  \rangle^{1/2}}{t} = \sqrt{\frac{2dD_t}{t}}
\end{equation}
For spherical droplets, the Stokes-Einstein relation provides a way to calculate the diffusion coefficient
\begin{equation}
	D_t = \frac{k_BT}{6\pi\eta R}
\end{equation}
Here \( \eta \) is the viscosity of the medium, and \( R \) is the radius of the droplet. Substituting \( D_t \) into the equation \ref{eq:vrms}, we get
\begin{equation}
	v_{rms} \propto \sqrt{\frac{1}{R}}
\end{equation}
Since for spherical droplets \( R \propto M_c^{1/3} \), the equation for \( v_{rms} \) becomes
\begin{equation}
	v_{rms} \propto M_c^{-1/6}
\end{equation}
This relation is verified in Fig.~\ref{fig:O_m_vrms}b from the simulation results (see caption). 

\section{Ballistic motion of clusters}
For ballistic motion, the MSD over time $t$ is given by
\begin{equation}
	\langle r^2(t) \rangle = v_{rms}^2 t^2
\end{equation}
At thermal equilibrium, the kinetic energy is related to the temperature $T$ through the equipartition theorem as
\begin{equation}
	\frac{1}{2} M_c v_{rms}^2 = \frac{3}{2} k_B T \quad ; \quad v_{rms} = \sqrt{\frac{3 k_B T}{M_c}}
\end{equation}
implying,
\begin{equation}
	v_{rms} \propto M_c^{-1/2}
\end{equation}
This relation is a fundamental characteristic of the thermal motion of particles, reflecting the balance between kinetic energy and mass.

\end{document}